\documentclass{osa-article}

\journal{oe}
\usepackage{graphicx}% Include figure files
\usepackage{dcolumn}% Align table columns on decimal point
\usepackage{bm}% bold math
\usepackage{amsmath}
\usepackage[utf8]{inputenc}
\usepackage[T1]{fontenc}
\usepackage[separate-uncertainty=true]{siunitx}

\begin{document}

\title{A high-power spectral beamsplitter for closely spaced frequencies}

\author{Ch. D. Marciniak\authormark{1,a}, A. Rischka\authormark{1,a,*}, R. N. Wolf\authormark{1} and M. J. Biercuk\authormark{1}}

\address{\authormark{1}School of Physics, The University of Sydney, NSW 2006, Australia, ARC Centre for Engineered Quantum Systems\\
\authormark{a}These authors contributed equally to this work}

\email{\authormark{*}alexander.rischka@sydney.edu.au} %% email address is required


\begin{abstract}
An experimental realization of a compact, high-power spectral beamsplitter for nearly equal frequencies and identical polarization based on two-beam interference in a free-space Mach-Zehnder interferometer is presented. We demonstrate the power- and cost-efficient generation and subsequent spatial separation of two laser tones from a single sum-frequency-generation stage using double-sideband suppressed-carrier modulation in the infrared, and beam splitting in the visible at high power.  The interferometer spectrally splits \SI{>98}{\percent} of the incident power when accounting for bulk absorption. The beamsplitter can be constructed identically for any power or spectral range required for which suitable optics are available.
\end{abstract}

%\ocis{(120.2440) Filters; (230.1360) Beam splitters; (230.7408) Wavelength filtering devices}

%%%%%%%%%%%%%%%%%%%%%%%%%%  body  %%%%%%%%%%%%%%%%%%%%%%%%%%

%\maketitle

\section{Introduction}

Precise control over the spectral and spatial properties of light is essential in various implementations of atomic physics, photonics, or quantum information processing based on trapped atoms and ions~\cite{goodwin2016resolved,roos2000experimental,eschner2003laser,sinclair2014spectral,schwartz2016generation}. In many of these applications multi-tone optical spectra with separations in the GHz range but otherwise identical properties are required. There are two major challenges arising when working with beams of closely spaced frequencies; Firstly, generation of such spectra can be cost inefficient and technically challenging, particularly if phase-coherence is required from independent sources; Secondly, splitting and recombination of spectral components at close frequency spacings can typically not be achieved with passive components whose discriminator slopes of a few nanometers width are many orders of magnitude too shallow for this task. Consequently, a range of different techniques have been developed for this task~\cite{abel2009faraday,haubrich2000lossless}, with a number of technical extensions for specific applications~\cite{bateman2010hansch,cooper2013stabilized}.

In this manuscript we aim to extend the existing body of work by presenting a simple, compact, and power-efficient system for the separation (or combination) of GHz-spaced tones of identical polarization utilizing a Mach-Zehnder beamsplitter (MZ-BS) from a single, phase-coherent beam.   We present details on the design, construction, and operation of the MZ-BS which operates at any power, optical frequency, or splitting for which suitable components are available. As a two-beam interferometer it does not require impedance or mode matching, performs best with convenient collimated input beams, and exhibits significantly relaxed length-stability requirements relative to alternative ring cavities~\cite{willke1998spatial,steinmetz2009fabry, palittapongarnpim2012note}.   Taken together these advantages enable superior performance in terms of alignment simplicity and reduced susceptibility to acoustic pickup. 

As a concrete example application, we demonstrate the generation and splitting of two phase-coherent laser tones separated by $\approx\SI{27.5}{\giga\hertz}$ for electromagnetically-induced transparency (EIT) cooling of 2D beryllium ion crystals in analog quantum simulation experiments~\cite{PenningPaper,EITExperiment}. The cooling laser beams are centered near \SI{313}{\nano\meter} and are generated by sum frequency generation (SFG) followed by resonantly-enhanced second harmonic generation (SHG).  Our approach reduces laser complexity by generating two narrowly-spaced seed-laser colors in the infrared, leveraging mature telecommunications-band waveguide-based modulation schemes.  This allows two tones near \SI{626}{\nano\meter} to be generated from a single SFG stage. We characterize the MZ-BS for beam separation in the visible near \SI{626}{\nano\meter} prior to frequency doubling, and demonstrate spectral splitting with $>98\%$ contrast ($ \approx \SI{500}{\milli\watt}$ per beam), consistent with theoretical predictions for optical losses in passive elements.  

\section{Mach-Zehnder beamsplitter}

\begin{figure}
\centering
\includegraphics[width=5.25in]{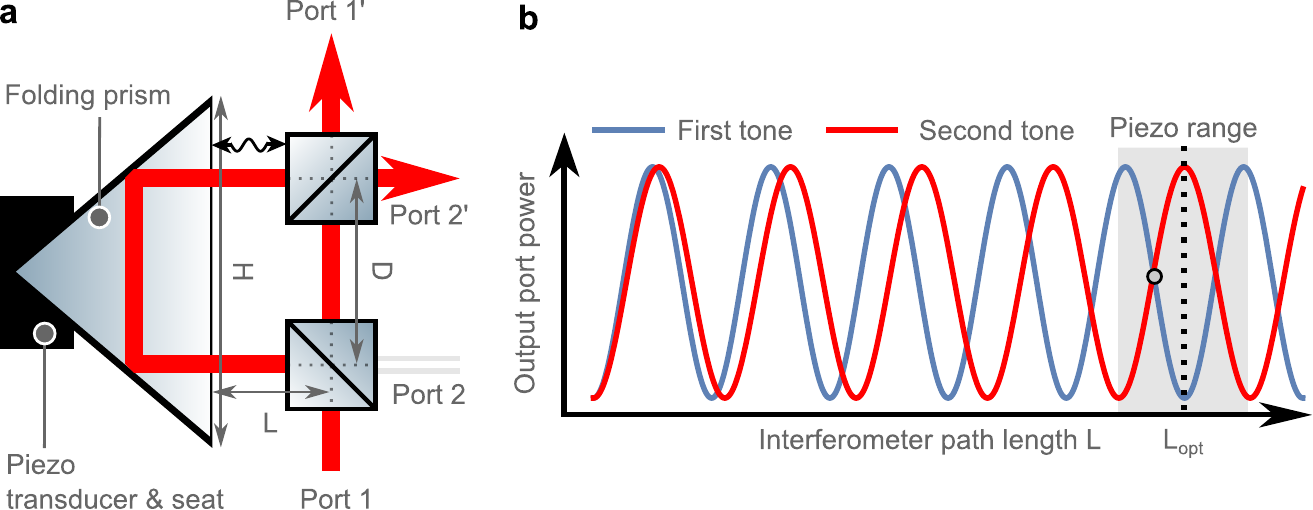}
\caption{\textbf{a} Schematic of the Mach-Zehnder interferometer design introduced here.  Optical elements include two discrete beamsplitter cubes, and a right angle folding prism in the long path, with pertinent dimensions indicated. The phase difference between the arms can be set by moving the prism (black wave arrow). \textbf{b} Schematic of output power at an output port as a function of interferometer dimension $L$ for two frequencies, where for presentation we pick a \SI{10}{\percent} frequency splitting.  Coarse tuning of $L$ determines whether both tones are in- or out-of-phase ($L_\text{opt}$ for optimal splitting) at the output, while fine tuning around the selected $L_\text{opt}$ via a piezo transducer (shaded area) is used to stabilize the interferometer at full splitting/combination (dashed line) rather than mixing (circle).}
\label{fig:MZI}
\end{figure}

The Mach-Zehnder beamsplitter (MZ-BS) is a Mach-Zehnder interferometer which is operated at a specific optical path length. In our implementation of the Mach-Zehnder interferometer, the input light is first split and then recombined on two 50:50 beamsplitters forming a two-path interferometer (Fig.~\ref{fig:MZI} \textbf{a}). The amount of light of a given frequency exiting one of the two output ports in such an interferometer is a function of the optical phase difference between the arms, which is frequency dependent.  For the task of separating two narrowly spaced spectral tones, the optical-path-length difference between the two arms is varied, here enabled by a movable folding prism in the long arm.  This parameter can be tuned to ensure constructive interference for one of the two input frequencies and destructive interference for the other on the same output port (Fig.~\ref{fig:MZI} \textbf{b}). Designing the MZ-BS thus requires finding the interferometer dimensions that achieve this~\cite{haubrich2000lossless} for a given frequency splitting. Given the geometric quantities defined in Fig.~\ref{fig:MZI} \textbf{a} we can find the optical path length difference between the arms as
\begin{equation}
    \Delta r = 2 L + n H - D,
\end{equation}
where $n$ denotes the refractive index of the prism material, and we measure both $L$ and $D$ from the center of the cubes. For separating two spectral tones on one input beam (or its time-reverse of combining two single-tone beam inputs into one two-tone output) we find the optical phase difference between the two paths for each of the two tones as
\begin{align}
    \delta_1 &= k_1\Delta r\\
    \delta_2 &= k_2\Delta r + \pi\\
    \Delta k &= k_2 - k_1 = 2\pi \Delta \nu / c,
\end{align}
where $k_i$ are the wave vectors of the two tones, $c$ is the speed of light in vacuum, and $\Delta \nu$ is the frequency splitting. Here the additional phase shift of $\pi$ between $\delta_1$ and $\delta_2$ comes from taking into account all phase shifts upon reflection from higher to lower optical density. We find for either configuration that the path difference for splitting/combination needs to be
\begin{equation}
    \Delta r \stackrel{!}{=} \frac{c}{\Delta \nu}\left(N_2 - N_1 - \frac{1}{2}\right),
\end{equation}
where $N_i$ are integer numbers. Generally, adjusting just one dimension of the interferometer, for example $L$, reduces complexity but is insufficient to guarantee complete interference in both frequencies at the same time; \emph{i.e.} for arbitrary $k_i$ we cannot enforce both $N_i$ to be integers. %However, practically there are a set of equally spaced values for the interferometer dimension $L$ that produce loss in contrast only negligibly small compared to the impact of experimental imperfections. We refer to these path lengths as $L_\text{opt}(\Delta\nu)$, since these lengths maximize the simultaneously achievable contrast, and they are given by
However, there are always interferometer dimensions $L$ that produce negligible loss in contrast compared to the ideal case, owing to the small relative frequency spacing. We refer to these interferometer dimensions as $L_\text{opt}(\Delta\nu)$, since they maximize the \emph{simultaneously achievable} contrast, with
\begin{equation}
    L_\text{opt} = \frac{1}{2}\left[D-n H + \frac{c}{\Delta\nu}\left(N_2-N_1 - \frac{1}{2}\right)\right].
\end{equation}
The rough length scale for the desired beam splitting, that is $L_\text{opt}$ at $\Delta\nu$, is set with a micrometer stage displacing the folding prism, while adjustment on interferometric length scales is provided by a piezoelectric transducer (Fig.~\ref{fig:MZI} \textbf{a}). In our application the relative frequency difference is $\mathcal{O}\left(10^{-4}\right)$ such that thousands of wave cycles need to be bridged by $L$ to tune the interferometer between in-phase and out-of-phase operation, far beyond the tuning range of the piezo. This means that $L$ typically has to be adjusted to within $\approx\SI{1}{\milli\meter}$ using the micrometer, in order to achieve high contrast. 

\begin{figure}
\includegraphics[width=5.25in]{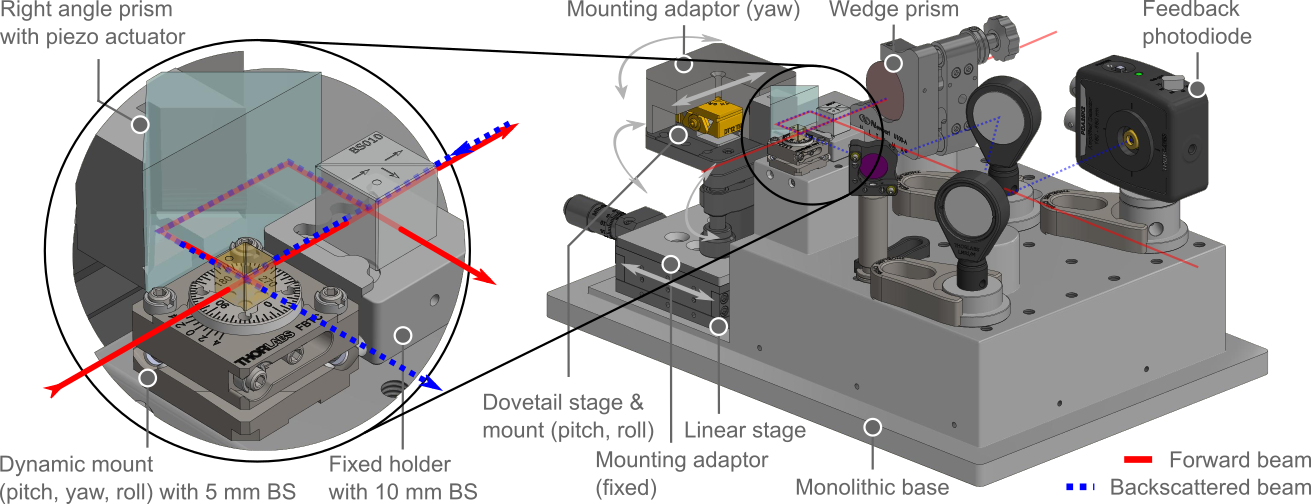}
\caption{Detail view of Mach-Zehnder beamsplitter with adjustment degrees of freedom and core parts indicated. The forward beam (red) is split, while the weak reflection from a wedge prism (blue) is used to derive an error signal to lock the interferometer length. Backscattered beam path shown is for interferometer length away from $L_\text{opt}$ with finite power on nominally dark port.}
\label{fig:machzehnder}
\end{figure}

Our implementation of a MZ-BS was constructed with a monolithic base having a footprint of $\approx 29\times\SI{17}{\centi\meter}$, as shown in Fig.~\ref{fig:machzehnder}, and is otherwise made from bulk components (Thorlabs and custom-machined). Roughly half of the surface area is taken by the interferometer, and the other half by a beam folding section which allows isolation of the feedback signal from other reflections in the optics. 
The interferometer itself is composed of a pair of non-polarizing beamsplitter cubes (Thorlabs BS007 and BS010, respectively sitting on Thorlabs FBTC and fixed holder) and a \SI{20}{\milli\meter} right-angle prism (Thorlabs PS908H-A on a custom seat) which is moved by a piezo transducer (Thorlabs PA4HKW) to allow adjustment of the optical-path-length difference of the interferometer. The cube centers are separated by \SI{21}{\milli\meter} and are positioned \SI{13.5}{\milli\meter} from the front face of the prism. We opt to use a right angle prism with adjusters instead of a fixed roof prism or corner cube due to the superior transmission for high power operation.

The prism assembly further features alignment degrees of freedom through its mounting: translation via Thorlabs DT12/M dovetail stage; pitch and roll through Thorlabs KM100B/M kinematic mount; and yaw through a custom mounting adapter.  The required translation used to adjust the total interferometer length towards $L_\text{opt}$ is accomplished using a Thorlabs MT1/M linear translation stage.  Similarly we provide pitch, yaw, and roll adjustment capabilities for the Thorlabs BS007 cube through the Thorlabs FBTC cube holder in order to compensate for machining and assembly imprecision. Bulk optics were selected in convenient sizes for ease of operation and reduced cost; the setup can be significantly compactified by employing a custom chevron prism, a beamsplitter-coated right angle prism~\cite{haubrich2000lossless}, and smaller detectors. The assembly features a removable lid to provide modest acoustic shielding~\cite{bies2017engineering} and laser safety.

% The interferometer length needs to be actively stabilized to maintain efficient operation. 
% We derive an error signal for interferometer-length stabilization by generating a back reflection at one output port from a circular wedge prism (Thorlabs PS812) and sending it back through the interferometer.  This back-reflected beam exits port 2 of the interferometer (Fig.~\ref{fig:MZI} \textbf{a} and Fig.~\ref{fig:setup})  where it is directed to a photodiode. Only one of the frequency components will be present at each output port if the interferometer length is near $L_\text{opt}$ and the piezo is set for complete interference. The backscattered, single-tone light will then see maximum destructive interference on the photodiode at port 2, minimizing the photodiode signal, which provides a dip-like error signal. Feedback is then implemented using a dither lock.

The interferometer length needs to be stabilized actively to maintain efficient operation regardless of whether the MZ-BS is used to split or combine beams. The main difference in operation of these two configurations is in how the error signal for this stabilization is derived. Derivation of the error signal in beam combination is straight-forward and uses the nominally dark output port 2' directly~\cite{haubrich2000lossless}. However, in beam splitting of two tones this is not possible. The total optical power at an output port is the sum of the two input tone powers, as shown in Fig.~\ref{fig:MZI} \textbf{b}. In the splitter configuration, the interferometer is operated near $L_\text{opt}$, where the sum of powers is constant regardless of piezo voltage, yielding no useful error signal on a power detector. Instead, we derive an error signal for interferometer-length stabilization by generating a back reflection from a circular wedge prism (Thorlabs PS812) and sending it back through the interferometer.  This back-reflected beam exits port 2 of the interferometer (Fig.~\ref{fig:MZI} \textbf{a} and Fig.~\ref{fig:setup})  where it is directed to a photodiode. Only one of the frequency components will be present at each output port if the interferometer length is near $L_\text{opt}$ and the piezo is set for complete interference. The backscattered, single-tone light will then see maximum destructive interference on the photodiode at port 2, minimizing the photodiode signal, which provides a dip-like error signal. Feedback is then implemented using a dither lock.

Alignment proceeds by optimizing the linear stage position while periodically scanning the path length using the piezo until the power on both output ports is constant irrespective of piezo voltage, thus fixing $L$ near $L_\text{opt}$. The back reflection from either output port is then aligned through the interferometer and onto the photodiode. The interference-fringe contrast on the photodiode will decrease if the back reflection is not  spectrally pure because the interferometer is adjusted away from all $L_\text{opt}$, or the power in both arms differs substantially. %If the interferometer is adjusted away from all $L_\text{opt}$,  the back reflection will not be spectrally pure, and the interference-fringe contrast on the photodiode will decrease.

\section{Experimental setup}

\begin{figure*}
\includegraphics[width=5.25in]{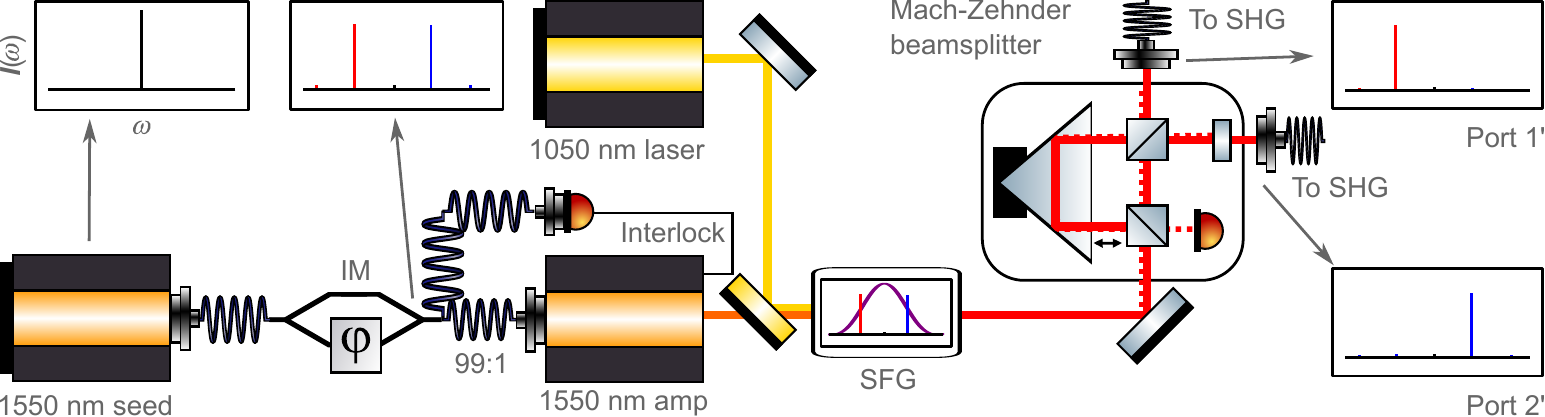}
\caption{Experimental setup schematic with intensity modulator (IM) operated for DSB-SC amplitude modulation between infrared pump laser seed and amplifier. The combined multi-tone infrared light passes a sum frequency generation (SFG) stage where both tones are converted to visible light. The Mach-Zehnder beamsplitter spatially separates the tones into two arms with near single-tone spectra, which are delivered to second harmonic generation (SHG) stages. Inset boxes show spectral intensity $I$ as a function of frequency $\omega$, SFG conversion efficiency indicated in purple.}
\label{fig:setup}
\end{figure*}

A schematic of the experimental setup used to efficiently generate and subsequently split the laser tones is shown in Fig.~\ref{fig:setup}. Our laser system is based on a cascading nonlinear frequency conversion scheme~\cite{Wilson2011} that begins with commercial fiber lasers and fiber amplifiers feeding into a single-pass sum frequency generation setup converting light to the visible.  High-power beam splitting is conducted in the red using the MZ-BS before visible light is fiber-coupled and fed to resonantly-enhanced second harmonic generation modules.

The two-tone laser spectrum is generated in the visible by modulation of a seed IR laser prior to amplification and SFG conversion to the red.  We insert a commercial waveguide intensity modulator (iXblue MXER-LN-20-PD-P-P-FA-FA-30dB) between the seed and amplifier of a custom \SI{1550}{\nano\meter} fiber laser (OrbitsLightwave INST-2500A-1549.60-5-PZ10B-TT60) providing direct fiber-based access to the seed light. The broadband modulator is driven by a microwave source at $\approx\SI{27.5/2}{\giga\hertz}$ and is biased to operate at the minimum of its transfer function to create only an upper (blue) and lower (red) sideband, while suppressing the carrier via double-sideband suppressed-carrier amplitude modulation (DSB-SC). This modulation scheme is advantageous as it minimizes power loss to the carrier or higher-order sidebands, and outperforms typical schemes involving a waveguide- or free space-electrooptic modulator and harmonic phase modulation. It is not necessary to stabilize this bias point, since the environmental conditions of our lab are well-controlled and the contrast in two-beam interferometers is a forgiving function of the ratio of intensities in the contributing arms.

A small portion of the light transmitted through the modulator is diverted via a 99:1 fiber tap to use as a safety interlock for the \SI{1550}{\nano\meter} laser amplifier, preventing unseeded operation. The modulated seed light after insertion losses and losses at the modulator is typically $\approx\SI{3}{\milli\watt}$, which is sufficient to drive the amplifier into saturation, fully compensating transmission losses. %, and only low-power components are necessary.
The amplified, modulated light is spatially overlapped in free space with a pure tone from a \SI{1050}{\nano\meter} fiber laser.  Both beams are then mixed inside a $\approx\SI{43}{\giga\hertz}$-bandwidth SFG stage with quasi-phase matching.  The temperature of the SFG stage is stabilized around \SI{180}{\degreeCelsius} to offset photorefractive damage, with a precision of \SI{\pm0.01}{\degreeCelsius} via a commercial oven controller (Covesion Ltd. PV40 oven with OC1 controller). 

The SFG stage produces two equal-strength tones at $\approx\SI{626}{\nano\meter}$, with total power $\approx\SI{1}{\watt}$.  The two spectral components are then split at the Mach-Zehnder beamsplitter, and each of the outputs is sent to individual resonantly-enhanced SHG stages to double to $\approx\SI{313}{\nano\meter}$ for EIT cooling.  The doubling process also doubles the frequency-spacing between the sidebands from $\approx\SI{27.5}{\giga\hertz}$ to $\approx\SI{55}{\giga\hertz}$, as required by the application.

\section{Beamsplitter performance}

% Characterize performance by bulk transmission (limited by absorption in cubes and prism), visibility (limited by alignment and cube performance asymmetry), and stability (limited by acoustics and locking performance).

We characterize the MZ-BS with regards to its ability to successfully separate the two spectral tones,  performance stability over time, and overall power efficiency.  We quantify the splitting efficiency by the achieved interferometric contrast $C$ through the interferometer. Low contrast means incomplete splitting, which directly translates to loss of power in the desired spectral component per arm. The contrast can be defined as
\begin{equation}
    C = \frac{I_\text{max}-I_\text{min}}{I_\text{max}+I_\text{min}} = \frac{2\sqrt{I_1I_2}}{I_1 + I_2},
\end{equation}
where $I_\text{max/min}$ are the maximum/minimum signal intensity as measured on a detector at the output port.  Here, $I_{1,2}$ are the intensities of light in each of the two interfering arms reaching the same port, assuming monochromatic light of identical polarization.  Measurements presented are for s-polarized input light, optimized for the passive optical elements in use. The predicted and measured contrast for s-polarized light assuming identical performance in both cubes is shown in Tab.~\ref{tab:visibilities}.
\begin{table}[h]
\centering
\renewcommand{\arraystretch}{1.5}
\setlength{\tabcolsep}{15pt}
\begin{tabular}{c|c c}\hline\hline
       Port  & $C_\text{theory}$ & $C_\text{measured}$ \\\hline
        1' & \SI{97.5 \pm 2}{\percent}&  \SI{98 \pm 1}{\percent}\\
        2' &\SI{99.9 \pm 2}{\percent}&  \SI{98 \pm 1}{\percent}  \\\hline\hline
\end{tabular}
\caption{Interferometer contrasts predicted from component specifications, and direct measurements, where the large theoretical errors stem from the wide component performance tolerance in the cubes.}
\label{tab:visibilities}
\end{table}

\begin{figure}[b]
    \centering
    \includegraphics[width=5.25in]{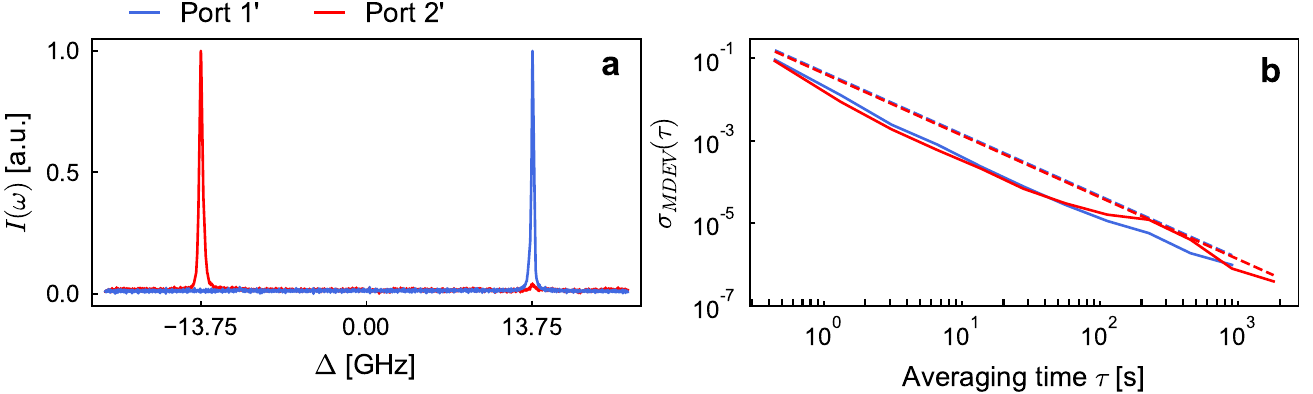}
    \caption{Mach-Zehnder beamsplitter performance. \textbf{a} Spectral content of the two output ports as a function of detuning from the carrier with the interferometer length stabilized as obtained by two scanning Fabry-Perot interferometers (FPIs). Plots are normalized due to different detector gain between the two FPIs. The carrier at no detuning is strongly suppressed, as are higher sidebands (not shown). \textbf{b} Modified Allan deviation of the ratio between transmitted and rejected sideband for both ports. The dashed lines are a guide to the eye with a slope of $\tau^{-3/2}$ consistent with an underlying flicker phase noise model in the fluctuations. The readout time resolution is limited to $T\approx\SI{0.5}{\second}$. }
    \label{fig:Stability}
\end{figure}

When the interferometer's length is stabilized we may measure the spectrum of the output light using a pair of Fabry-Perot interferometers (Thorlabs SA200-5B) aligned to each of the output ports, as shown in Fig.~\ref{fig:Stability} \textbf{a}. We characterize the stability of the interferometer by measuring the ratio of the transmitted and rejected sidebands for each output port, and then calculating the modified Allan deviation of this ratio as a function of averaging time $\tau$ (Fig.~\ref{fig:Stability} \textbf{b}). We elect to use this measure over the more common (overlapping) Allan deviation because it can distinguish between flicker- and white-phase noise averaging through the power-law scaling in $\tau$, while the other two can not. The $\tau^{-3/2}$ scaling indicates an underlying flicker-phase noise fluctuation in the ratio~\cite{rutman1978characterization}. 

Overall, bulk transmission through the device is measured at \SI{81}{\percent} limited by the cube and prism absorption of $\approx\SI{6}{\percent}$ each; this number can be easily improved by using lower-absorption components. Observed performance depends on input polarization state mainly due to performance variation in the beamsplitter coatings.

\section{Conclusion}

In this work we presented and characterized a Mach-Zehnder beamsplitter to separate two closely spaced spectral components imposed on a single beam. With our choice of optics we found that $>\SI{98}{\percent}$ of the power in each spectral component exits on the correct port, and that fluctuations in the splitting ratio of the actively length-stabilized assembly scale in time like a flicker-phase-noise model for measurements up to \SI{10}{\minute}. This phase noise is likely introduced within the electronics of the feedback chain, and may pose a limitation if long-timescale stability in the intensity or spectral purity is required. In our application, spectral purity is guaranteed by the resonance condition of the SHG units, and low frequency intensity fluctuations are easily removed by downstream intensity stabilization.  In our implementation both interferometer footprint and transmission performance are limited by design choices optimizing convenience and cost, and can be improved by utilizing more specialized components and assemblies.

In our atomic-physics application, the setup described here saves considerably in cost and overhead compared with producing two beams from separate nonlinear frequency conversion and phase-locked source lasers.  This is because it requires no GHz-scale detection or lock electronics, and only employs a single laser source. Power- and cost-efficient generation of two tones from a single source is available as turn-key solutions from near DC to sideband separations of $\Delta\nu\approx\SI{50}{\giga\hertz}$ prior to SFG by utilizing commercial telecommunications-band waveguide intensity modulators prior to a saturated amplifier and double-sideband suppressed-carrier amplitude modulation. At increasingly large frequency splittings other considerations like the conversion bandwidth of the nonlinear medium for SFG are likely to become limiting factors in available power, rather than the MZ-BS. Two-tone light may be physically split by the Mach-Zehnder beamsplitter at any power and carrier frequency for which suitable optics are available. In the future we will use this system to generate the required phase-coherent beams for EIT cooling of a 2D crystal beryllium ions in a Penning trap close to their motional groundstate.  We are hopeful that the simplifications and cost-reductions afforded by this stable MZ-BS system will prove useful in a variety of atomic and optical physics experiments, where spatially splitting or combining laser beams with closely spaced frequencies is required.

\section{Funding}
Australian Research Council Centre of Excellence for Engineered Quantum Systems (EQUS) (CE170100009). R.N.W. acknowledges support by the Australian Research Council under the Discovery Early Career Researcher Award scheme (DE190101137). 

\section{Disclosures}
The authors declare that there are no conflicts of interest related to this article.

\bibliography{references}% Produces the bibliography via BibTeX.

\end{document}